\documentclass[prb,groupaddress,twocolumn,longbibliography]{revtex4-2}
\usepackage{CJK}
\usepackage{graphicx,color,dcolumn,bm}
\usepackage[colorlinks=true,breaklinks=true,allcolors=blue]{hyperref}
\usepackage{natbib}

\begin{document}
\begin{CJK*}{UTF8}{bsmi}
\title{First Principle Study for Optical Properties of TMDC/Graphene Heterostructures}
\author{Cheng-Hsien Yang (楊承憲)}
\email{d104064006@mail.nchu.edu.tw}
\author{Shu-Tong Chang}
\affiliation{Department of Electrical Engineering, National Chung Hsing University, Taichung 40227, Taiwan}


\begin{abstract}
The transition-metal dichalcogenide (TMDC) in the family of $\mathrm{MX}_2$ ($\mathrm{M}=\mathrm{Mo},\mathrm{W}$; $\mathrm{X}=\mathrm{S},\mathrm{Se}$) and the {graphene} ($\mathrm{Gr}$) monolayer are an atomically thin semiconductor and a semimetal, respectively. The monolayer $\mathrm{MX}_2$ has been discovered as a new class of semiconductors for electronics and optoelectronics applications. Because of the hexagonal lattice structure of both materials, $\mathrm{MX}_2$ and $\mathrm{Gr}$ are often combined with each other to generate van der Waals heterostructures. Here, the $\mathrm{MX}_2/\mathrm{Gr}$ heterostructures are investigated theoretically based on density functional theory (DFT). The electronic structure and the optical properties of four different $\mathrm{MX}_2/\mathrm{Gr}$ heterostructures are computed. We systematically compare these $\mathrm{MX}_2/\mathrm{Gr}$ heterostructures for their complex permittivity, absorption coefficient, reflectivity and refractive index.
\end{abstract}

\maketitle
\end{CJK*}
\section{Introduction}
Materials in confined geometries are hosts to novel phenomena. For example, in one dimension (1D), the fractional charge excitation or the soliton is found in the electronic properties of the organic polyacetylene~\cite{SSH1979}, and the Haldane conjecture~\cite{Haldane}, which corresponds to the symmetry-protected topological phase, is experimentally found in the spin-1 magnetic chain, e.g., CsNiCl$_3$, [Ni(HF$_2$)(3-Clpyradine)$_4$]BF$_4$, Ni(C$_2$H$_8$N$_2$)$_2$NO$_2$(ClO$_4$) and Y$_2$BaNiO$_5$ \cite{Y2BaNiO5,CsNiCl3,NENP,Tzeng2012,Tzeng2017,high-pressure}. Quantum states conversion through entanglement manipulations~\cite{Tzeng2016} as well as new types of quantum phase transitions with a negative central charge~\cite{Tu2022} are among the recently investigated 1D systems.

Examples of novel physical phenomena in two dimension (2D) include quantum Hall systems in semiconductors \cite{QHE}, the supersolid phase of atomic systems in the optical lattice \cite{Ng_2010,Chen_2017} and Dirac electrons in graphene \cite{geim2009graphene}. Graphene (Gr) is one of the most widely investigated 2D materials; it possesses a high mobility \cite{bolotin2008ultrahigh, c-04, c-08, c-32, c-33} and is driven by Dirac cone dispersions near the Fermi level at the K-point and the K'-point in the hexagonal Brillouin zone\cite{RMP2009}. However, the gapless nature of Gr limits its applications in the semiconductor industry. Therefore, finding new 2D materials with a large band gap is an interesting topic. Note that a new allotropy of carbon, the 2D biphenylene monolayer, was discovered very recently \cite{Biphenylene:science, Biphenylene:srep, Biphenylene:JPCM, referee2}.

In addition to the carbon, a number of 2D materials with an energy band gap have been synthesized, including the quantum spin Hall insulator HgTe \cite{SCZhang:science-HgTe,HgTe}, free-standing silicene \cite{Silicene:PRL} and the transition-metal dichalcogenide (TMDC) family. Monolayer TMDCs~\cite{TMDC:science2011,TayRong-PRB2015,TayRong-Natcomm2017,TayRong-Natnano2014, CHYang2022, c-02, c-03, MoS2:band-PRB,c-07, c-21, c-31, c-35}, such as molybdenum disulfide (MoS$_2$) \cite{TayRong-SciRep2014,c-15}, molybdenum diselenide (MoSe$_2$) \cite{MoSe2:PRL2021}, tungsten disulfide (WS$_2$) \cite{WS2} and tungsten diselenide (WSe$_2$) \cite{c-24,RoYa2019PRB}, have been found to be direct gap semiconductors \cite{c-14} and have emerged as new optically active materials for novel device applications. The TMDCs in the family of MX$_2$, where $\mathrm{M}=\mathrm{Mo}, \mathrm{W}$ and $\mathrm{X}=\mathrm{S}, \mathrm{Se}$, have a rather large energy band gap compared to HgTe and silicene; thus, these materials are more versatile as candidates for thin, flexible device applications and useful for a variety of applications including transistors \cite{MoS2:transistor}, lubrication \cite{science.1184167}, lithium-ion batteries \cite{C1CC10631G} and thermoelectric devices \cite{CJP:thermo}. Other TMDCs, e.g., PbTaSe$_2$, Mo$_x$W$_{1-x}$Te$_2$ and NiTe$_2$, can exhibit more unusual physical properties such as topological superconductivity and Dirac and Weyl semimetal physics \cite{TayRong-PRB2016,TayRong-Natcomm2016a,TayRong-Natcomm2016b,Tzeng2020,CJP:newTI,NiTe2}. Two-dimensional hydrogenated NiTe$_2$, PdS$_2$, PdSe$_2$, and PtSe$_2$ monolayers also exhibit a quantum anomalous Hall effect~\cite{CJP:QAHE}.

One of the approaches used to broaden the application of 2D materials is to form a heterostructure in which similar lattice structures for different monolayers are usually required, e.g., the hexagonal lattices of TMDC, Gr, boron nitride~\cite{c-01}, and boron phosphide (BP) \cite{referee1}. It has been suggested that MX$_2$/BP heterostructures have a good ability for optical absorption \cite{referee1}. On the other hand, the MX$_2$/Gr of the type-I van der Waals heterostructures have attracted much attention for their electronic properties. Research on MoS$_2$/Gr \cite{c-16,c-27,c-09}, MoSe$_2$/Gr \cite{c-05,c-19,c-20}, WS$_2$/Gr \cite{c-22} and WSe$_2$/Gr \cite{c-25,c-26} has revealed that MX$_2$/Gr is favorable for electronics applications and field-effect transistors. However, there are relatively few studies on its optical properties \cite{5c,14c,21c}; therefore, comprehensive research on the optical properties of MX$_2$/Gr is desired.

In this communication, we theoretically investigate the optical properties of MX$_2$/Gr heterostructures. By means of a density functional theory simulation, we systematically compare the energy band structures, the density of states, complex permittivity, absorption coefficients, reflectivity, and refraction indexes.

\section{Materials and Methods}
\begin{figure*}[t]
\centering
\includegraphics[width=5in]{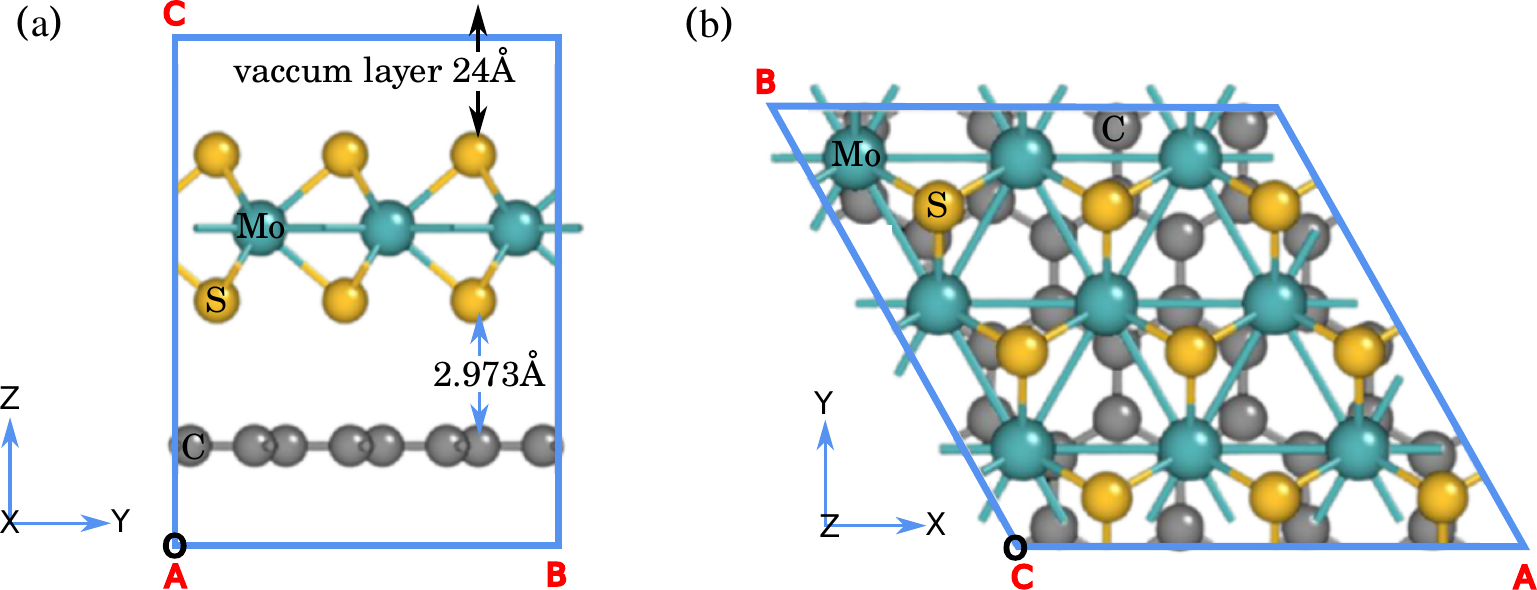}
\caption{The stacking of the $3\times3$ MoS$_2$ supercell and the $4\times4$ {Gr} supercell. (a)~The side view. (b)~The top view. For the other MX$_2$/Gr, the atom--atom distances are listed in Table~\ref{tab:optimization}.}\label{fig:stacking}
\end{figure*}

\begin{table}[b]
\caption{The lattice mismatch ratios and the biding energies of TMDCs/graphene with $3\times3$ supercell of $\mathrm{MX}_2$ and $4\times4$ supercell of graphene.}\label{tab:mismatch}
\begin{ruledtabular}
\begin{tabular}{ccccc}
$\mathrm{MX}_2$/Gr & $\mathrm{MoS}_2$/Gr & $\mathrm{MoSe}_2$/Gr & $\mathrm{WS}_2$/Gr & $\mathrm{WSe}_2$/Gr \\ \hline
mismatch ratio & 1.83\% & 0.10\% & 2.04\% & 0.01\%  \\
$E_b$ & $-194$~meV & $-146$~meV & $-178$~meV & $-246$~meV \\
\end{tabular}
\end{ruledtabular}
\end{table}

The first-principles calculations are based on the local-density approximation (LDA) {proposed by Kohn and Sham \cite{Kohn-Sham}, which approximates the total energy of the multielectron system.} The simulations were implemented using the software package \textsc{Nanodcal} and its accompanying software packages \textsc{DeviceStudio} and \textsc{OpticCal}~ \cite{nanodcal:1,nanodcal:2}. A plane wave basis was set with a cutoff energy of $80$ Hartree and a $5\times5\times1$ $\Gamma$-centered $k$-points grid. The atomic structures were relaxed until the force was smaller than $0.03$ eV/\AA{} and the total energy convergence criterion was set as $10^{-4}$~eV. In order to avoid interactions in the vertical direction between neighboring slabs, a vacuum layer of 24~\AA{} was added between different slabs.

\begin{table}[b] 
\caption{The optimized atom-atom distances as well as the $\mathrm{MX}_2$ layer -- Gr layer distances.}\label{tab:optimization}
\begin{ruledtabular}
\begin{tabular}{cccccc}
$\mathrm{MoS}_2$/Gr & Mo-S & Mo-C & S-C & C-C & $\mathrm{MoS}_2$-Gr\\ \hline
& 2.404\AA{} & 4.826\AA{} & 3.089\AA{} & 1.600\AA{} & 2.973\AA{} \\ \hline
$\mathrm{MoSe}_2$/Gr & Mo-Se & Mo-C & Se-C & C-C & $\mathrm{MoSe}_2$-Gr\\ \hline
& 2.349\AA{} & 4.920\AA{} & 3.270\AA{} & 1.626\AA{} & 3.161\AA{} \\ \hline
$\mathrm{WS}_2$/Gr & W-S & W-C & S-C & C-C & $\mathrm{WS}_2$-Gr\\ \hline
& 2.379\AA{} & 4.774\AA{} & 3.078\AA{} & 1.596\AA{} & 2.954\AA{} \\ \hline
$\mathrm{WSe}_2$/Gr & W-Se & W-C & Se-C & C-C & $\mathrm{WSe}_2$-Gr\\ \hline
& 2.354\AA{} & 4.940\AA{} & 3.286\AA{} & 1.625\AA{} & 3.178\AA{}
\end{tabular}
\end{ruledtabular}
\end{table} 
 
The lattice constants of the MoS$_2$, MoSe$_2$, WS$_2$, WSe$_2$ and Gr monolayers were 3.166~\AA{}, 3.288~\AA{}, 3.153~\AA{}, 3.282~\AA{} and 2.47~\AA{}, respectively\cite{CHYang2022,c-05}. For all the MX$_2$/Gr heterostructures, lattice mismatch ratios less than 5\% were achieved by choosing a $3\times3$ MX$_2$ supercell and a $4\times4$ Gr supercell, as shown in Figure \ref{fig:stacking}. The details of the lattice mismatch ratios are listed in Table \ref{tab:mismatch}. After the structure optimization, the lattice contents were 9.668 \AA{}, 9.854 \AA{}, 9.648 \AA{} and 9.845 \AA{} for MoS$_2$/Gr, MoSe$_2$/Gr, WS$_2$/Gr and WSe$_2$/Gr, respectively. The details of the atom--atom distances as well as the $\mathrm{MX}_2$ layer--Gr layer distances are listed in Table \ref{tab:optimization}. 
The total number of atoms in the simulations was 59 for MX$_2$/Gr, 27 for the $3\times3$ monolayer of MX$_2$, and 32 for the $4\times4$ monolayer of Gr.

The binding energies $E_b$ are calculated by the energy difference between the heterostructures and the monolayers:
\begin{equation}
E_b=E_{\mathrm{MX}_2/\mathrm{Gr}}-E_{\mathrm{MX}_2}-E_{\mathrm{Gr}},
\end{equation}
where $E_{\mathrm{MX}_2/\mathrm{Gr}}$, $E_{\mathrm{MX}_2}$ and $E_\mathrm{Gr}$ are the total energies of the heterostructures, isolated MX$_2$ and Gr, respectively. It is interesting to compare 
	other heterostructures, such as MX$_2$/BP~ \cite{referee1}, as regards binding energies. The interface binding energies of the most stable configurations of MoS$_2$/BP, MoSe$_2$/BP, WS$_2$/BP and WSe$_2$/BP are $-196$~meV, $-130$~meV, $-201$~meV and $-141$~meV, respectively,  \cite{referee1}. 
As listed in Table \ref{tab:mismatch}, all the binding energies for the MX$_2$/Gr heterostructures in this study are negative, and the values are close to MX$_2$/BP, demonstrating that the structural stability of MX$_2$/Gr is similar to that of MX$_2$/BP.

\begin{figure*}[t]
\centering
\includegraphics[width=5.4in]{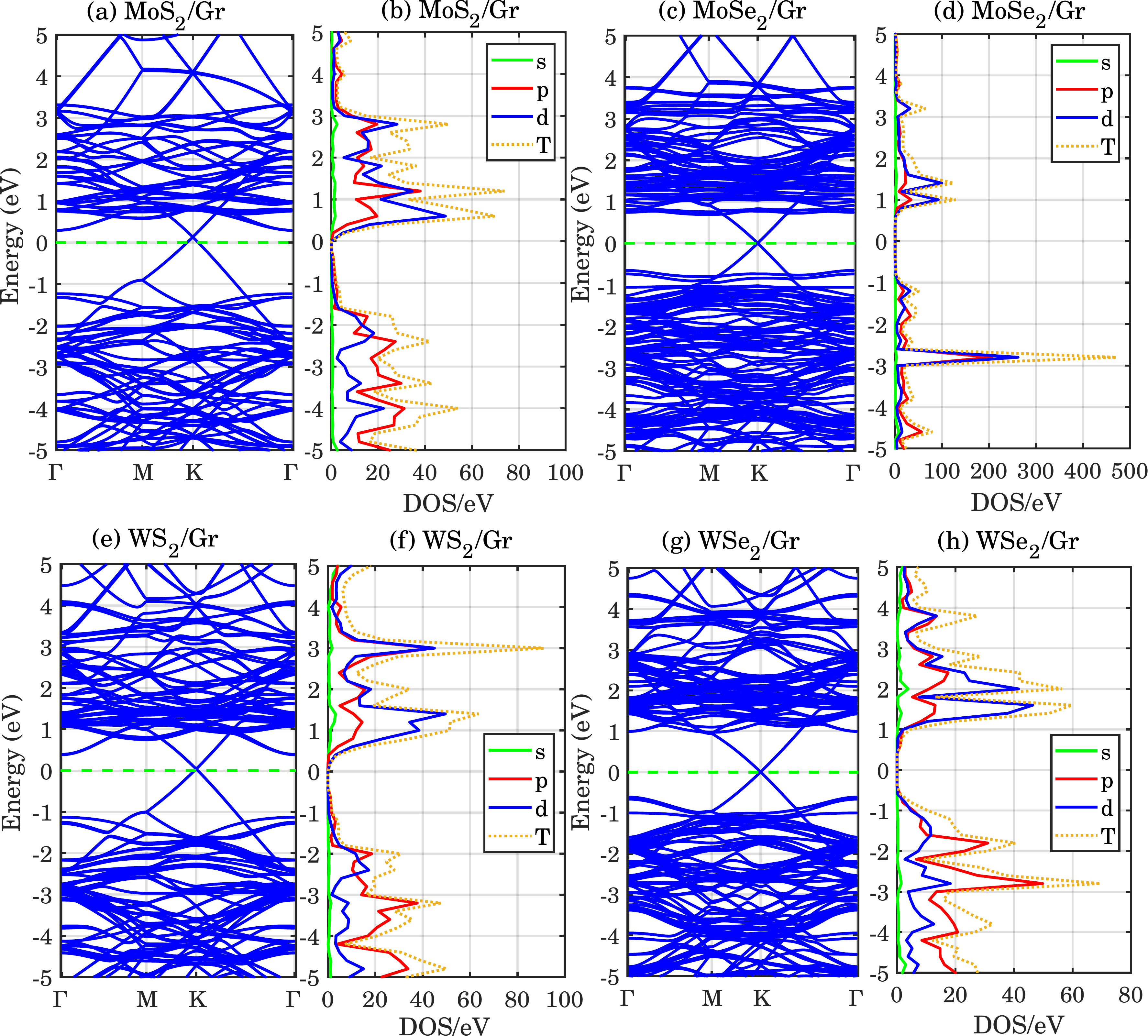}
\caption{The energy band structures and the DOS for the MX$_2$/Gr van der Waals heterostructures. The partial DOS values are labeled with s, p, or d, corresponding to the orbital contributions, and the total DOS is labeled with T. (a,b) MoS$_2$, (c,d) MoSe$_2$, (e,f) WS$_2$, (g,h) WSe$_2$.}\label{fig:MX2-Gr-bdt}
\end{figure*}

\section{Results}
\subsection{Band structures}
The direct band gaps of the MoS$_2$, MoSe$_2$, WS$_2$ and WSe$_2$ monolayers were $1.82$~eV, $1.56$~eV, $1.95$~eV and $1.64$~eV, respectively, \cite{c-07}, where both the conduction band minimum (CBM) and the valence band maximum (VBM) were at the K-point in the Brillouin zone. The Dirac point of the {Gr} was located at the K-point and pinned to the Fermi level. All the results of the energy band structures for the MX$_2$ monolayers are consistent with the existing literature \cite{MoS2:band-PRB,c-08}. Now we report the energy band structures and the electronic density of states (DOS) for the four MX$_2$/Gr, as shown in Figure \ref{fig:MX2-Gr-bdt}. We found that the band structure of MX$_2$/Gr could basically be regarded as the overlap of the band structures of the MX$_2$ and Gr monolayers. Due to van der Waals interactions, the Dirac point opens a small gap of $8.3$~meV, $7.9$~meV, $8.86$~meV and $3.53$~meV for MoS$_2$/Gr, MoSe$_2$/Gr, WS$_2$/Gr and WSe$_2$/Gr, respectively.
The total and partial DOS values of MX$_2$/Gr are also shown in Figure \ref{fig:MX2-Gr-bdt}. The partial DOS is the relative contribution of a particular orbital to the total DOS. The d-orbital contributions mainly come from the W or Mo atoms, and the p-orbital contributions come from S or Se, and C atoms. Contributions from the s-orbital are minimal. These results reflect a fairly weak interfacial coupling at the interface between MX$_2$ and Gr.

The Schottky contacts are formed between the semiconducting MX$_2$ and the metallic Gr. The Schottky barrier is one of the most important characteristics of a semiconductor--metal junction and dominates the transport properties. Based on the Schottky--Mott model~\cite{PhysRev.71.717,c-11,c-12}, at the interface of the metal and semiconductor, an n-type Schottky barrier height (SBH) $\Phi_{Bn}$ is defined as the energy difference between the Fermi level $E_F$ and the conduction band minimum $E_C$, i.e., $\Phi_{Bn}=E_C-E_F$. Similarly, the p-type SBH $\Phi_{Bp}$ is defined as the energy difference between $E_F$ and the valence band maximum $E_V$, i.e., $\Phi_{Bp}=E_F-E_V$. The n/p-type Schottky contacts are classified by the smaller SBH. The SBH are $\Phi_{Bn}=0.29~\mathrm{eV}$ and $\Phi_{Bp}=1.23~\mathrm{eV}$ for MoS$_2$/Gr, $\Phi_{Bn}=0.75~ \mathrm{eV}$ and $\Phi_{Bp}=0.65~\mathrm{eV}$ for MoSe$_2$/Gr, $\Phi_{Bn}=0.39~\mathrm{eV}$ and $\Phi_{Bp}=1.13~\mathrm{eV}$ for WS$_2$/Gr, and $\Phi_{Bn}=1.00~\mathrm{eV}$ and $\Phi_{Bp}=0.61~\mathrm{eV}$ for WSe$_2$/Gr, respectively. Therefore, MoS$_2$/Gr, MoSe$_2$/Gr, WS$_2$/Gr and WSe$_2$/Gr are the n-type, p-type, n-type and p-type Schottky contacts, respectively\cite{c-19,c-22,c-25}.

\subsection{Optical properties}
\begin{figure*}[t]
\centering
\includegraphics[width=5.4in]{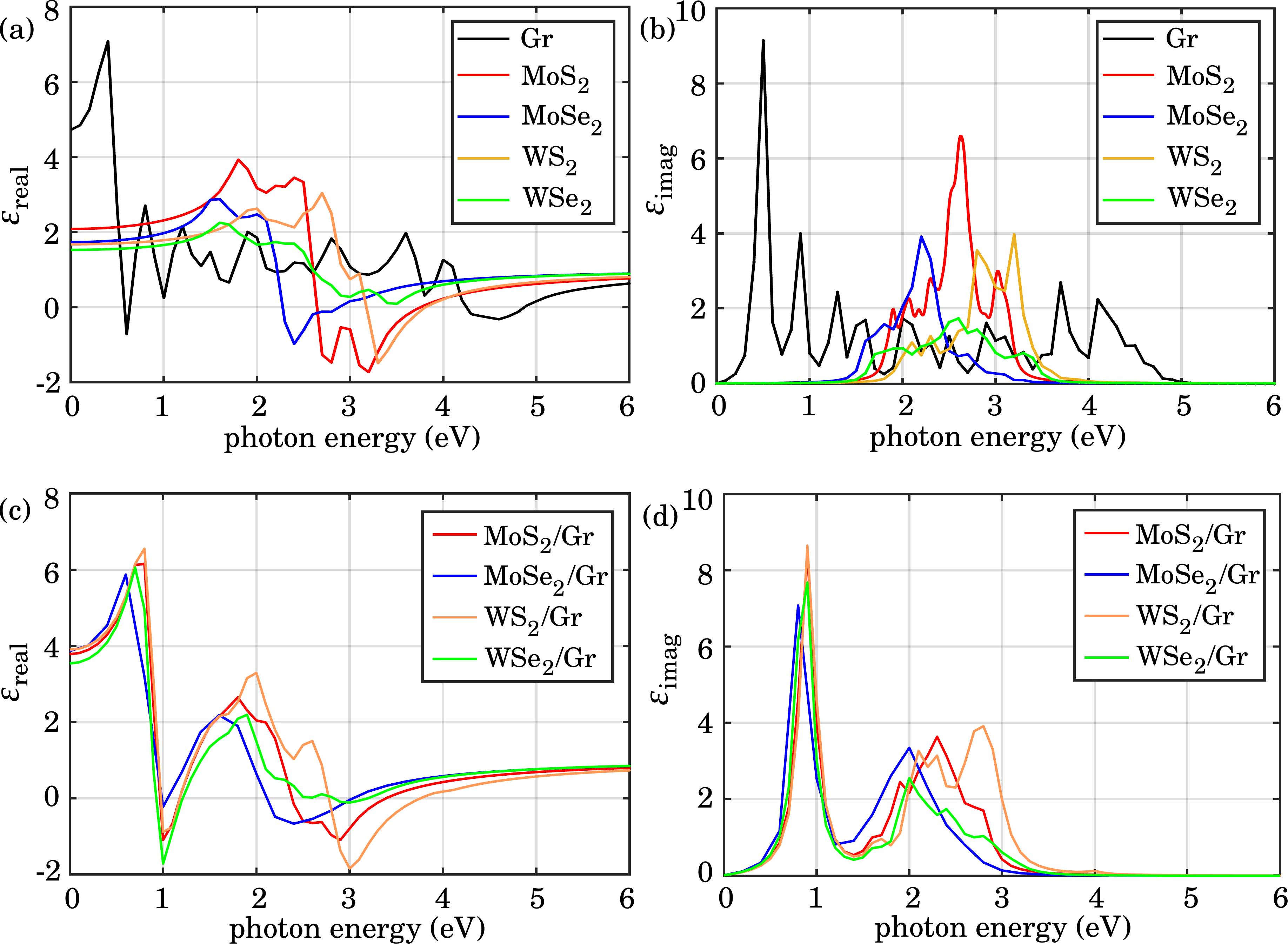}
\caption{The complex permittivities for MX$_2$/Gr heterostructures and MX$_2$ and Gr monolayers as functions of incident photon energy. (a),(c) The real part of the permittivity $\varepsilon_\mathrm{real}$. (b),(d) The imaginary part of the permittivity $\varepsilon_\mathrm{imag}$.}\label{fig3}
\end{figure*}

In order to understand the optical properties, the complex permittivity or the so-called dielectric function was computed under the long-wave approximation, i.e., $\vec q=0$.
The complex permittivity $\varepsilon(\omega)=\varepsilon_\mathrm{real}(\omega)+\varepsilon_\mathrm{imag}(\omega)$ as a function of incident photon energy is \cite{grosso2013solid}
\begin{equation}
\varepsilon(\omega)=1+\frac{2e^2}{\varepsilon_0m^2}\frac{1}{V}
\sum_{\alpha\beta}\frac{|\langle\psi_\beta|\hat{e}\cdot\vec p|\psi_\alpha\rangle|^2}{(E_\beta-E_\alpha)^2/\hbar^2}
\frac{f(E_\alpha)-f(E_\beta)}{E_\beta-E_\alpha-\hbar\omega-i\eta}
\end{equation}
where $e$ is the electron charge, $\varepsilon_0$ is the vacuum permittivity, $m$ is the electron mass, $\hat{e}$ is the direction of the vector potential, $\vec p$ is the momentum operator, $\hbar\omega$ is the photon energy, and $f(E_\alpha)$ and $f(E_\beta)$ are the Fermi--Dirac distribution functions.
Since $\varepsilon=n_c^2$, i.e., $\varepsilon_\mathrm{real}+i\varepsilon_\mathrm{imag}=(n+i\kappa)^2$, the refraction index $n$ and the extinction coefficient $\kappa$ are obtained:
\begin{equation}
        n^2=\frac{|\varepsilon|+\varepsilon_\mathrm{real}}{2},
\end{equation}
\begin{equation}
        \kappa^2=\frac{|\varepsilon|-\varepsilon_\mathrm{real}}{2}.
\end{equation}

The real part $\varepsilon_\mathrm{real}$ is caused by various kinds of displacement polarization inside the material and represents the energy storage term of the material. The imaginary part $\varepsilon_\mathrm{imag}$ is related to the absorption of the material, including gain and loss. Therefore, the permittivity must be real in the absence of the incident photon energy, i.e., $\varepsilon(\omega=0)=\varepsilon_\mathrm{real}$. It is expected that the smaller the energy gap of a material, the larger its $\varepsilon(0)$. On the other hand, since the DOS only appears in a finite range of energy in the numerical simulation, electron transitions by the absorption of photons do not occur if the photon energy is too large. Therefore, without photon absorption the computed $\varepsilon_\mathrm{imag}$ is close to zero in the large limit of the incident photon energy, and the permittivity approaches a constant real value.

In Figure \ref{fig3}, we show the real part and the imaginary part of the complex permittivity for MX$_2$/Gr heterostructures and MX$_2$ and Gr monolayers.
Since the energies corresponding to the peak positions of the DOS in the conduction and valence bands of the MX$_2$/Gr were smaller than those of MX$_2$, the peak positions of the real part $\varepsilon_\mathrm{real}(\omega)$ of MX$_2$/Gr in Fig.~\ref{fig3}(c) had a red shift compared with the MX$_2$ in Fig.~\ref{fig3}(a). The highest peak corresponding energy of the imaginary part $\varepsilon_\mathrm{imag}(\omega)$ of the MX$_2$/Gr in Fig.~\ref{fig3}(d) was about 0.8~eV to 0.9~eV, which corresponds to the position where the real part decreases the fastest.

\begin{figure*}[t]
\centering
\includegraphics[width=6.7in]{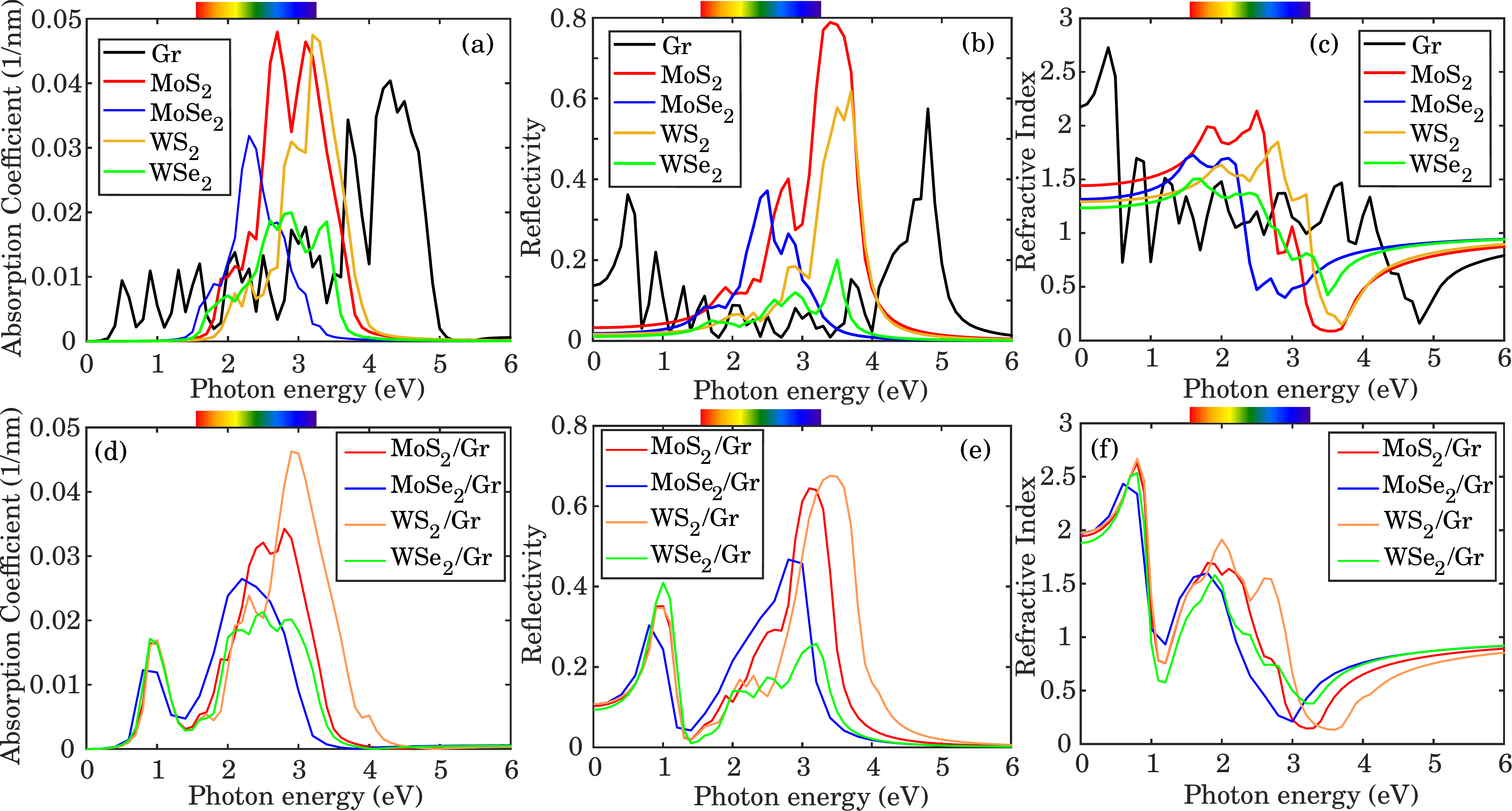}
\caption{The comparison of MX$_2$/Gr heterostructures and MX$_2$ and Gr monolayers according to their (a),(d) absorption coefficient $\alpha$, (b),(e) reflectivity $R$ and (c),(f) refractive index $n$. The rainbow bars represent the energy range of visible light, from $1.59$~eV to $3.26$~eV.}\label{fig4}
\end{figure*}

After obtaining the dielectric function, the absorption coefficient $\alpha$ is computed by
\begin{equation}
\alpha=\frac{2\omega\kappa}{c},
\end{equation}
and the reflectivity $R$ is
\begin{equation}
R=\frac{(n-1)^2+\kappa^2}{(n+1)^2+\kappa^2}.
\end{equation}

Figure \ref{fig4} compares MX$_2$/Gr, MX$_2$ and Gr for their absorption coefficient~$\alpha$, reflectivity~$R$ and refractive index~$n$. Since MX$_2$/Gr has a smaller optical band gap compared with the MX$_2$ monolayer, MX$_2$/Gr has a wider range of light absorption, from $0.6$~eV to $4$~eV. As shown in Fig.~\ref{fig4}(b),(e), MX$_2$/Gr had a higher reflectivity than MX$_2$ in the infrared area ($\hbar\omega~{<}~1.2~\mathrm{eV}$), and had a higher reflectivity than Gr in the visible light area. We compare the real part of the permittivity of monolayer MX$_2$ in Fig.~\ref{fig3}(a) with the refractive index of monolayer MX$_2$ in Fig.~\ref{fig4}(c), and compare the real part of the permittivity of the MX$_2$/Gr heterostructure in Fig.~\ref{fig3}(c) with that of MX$_2$/Gr in Fig.~\ref{fig4}(f). It was found that the changing trends from Fig.~\ref{fig3}(a) to Fig.~\ref{fig4}(c) are similar to those from Fig.~\ref{fig3}(c) to Fig.~\ref{fig4}(f). This means that the real part of the dielectric constant dominates the effect of the refractive index. The above simulation results suggest that MX$_2$/Gr heterostructures are good candidate materials for optical applications.

\section{Discussion}
Two-dimensional heterostructures based on TMDCs exhibit the enhancement of electrical and optoelectrical properties, which are promising for next-generation optoelectronics devices. We systematically computed the complex permittivity $\varepsilon(\omega)$, absorption coefficient $\alpha(\omega)$, reflectivity $R(\omega)$ and refractive index $n(\omega)$ for MX$_2$/Gr heterostructures, where M = Mo,~W; and X = S,~Se. Our results {qualitatively} agree with those from previous studies on MoS$_2$/Gr~\cite{5c} and WSe$_2$/Gr \cite{14c}, where red shifts in the $\alpha(\omega)$, $R(\omega)$ and $n(\omega)$ were found compared with MoS$_2$ and WSe$_2$ monolayers. We extended the investigations to other MX$_2$/Gr heterostructures, and found qualitatively similar behavior in their optical properties.

It is worth comparing our MX$_2$/Gr results with the recent simulation results on the MX$_2$/BP in terms of absorption abilities \cite{referee1}. Although different types of van der Waals heterostructures were found in MX$_2$/BP (type-I for MoS$_2$/BP and WS$_2$/BP; type-II for MoSe$_2$/BP and WSe$_2$/BP), all the materials of MX$_2$/BP have excellent absorption abilities in the infrared and visible light range, i.e., $\alpha(\omega)\approx0.01~\mathrm{nm}^{-1}$ to $0.05~\mathrm{nm}^{-1}$ for the wavelength $400~\mathrm{nm}\leq\lambda\leq1200~\mathrm{nm}$~ \cite{referee1}. In our study, all the MX$_2$/Gr were type-I heterostructures, and the values of the absorption coefficients were in the same range compared with MX$_2$/BP in the infrared and visible light range. Therefore, MX$_2$/Gr can be utilized as alternative materials for the applications of solar optoelectronics devices.

\begin{acknowledgements}
CHY is grateful to Dr. Yu-Chin Tzeng for his invaluable discussions and helps.
\end{acknowledgements}

%

\end{document}